\documentclass[seceq,twocolumn]{jpsj2}
\begin{document}
\setlength{\textwidth}{504pt}
\setlength{\columnsep}{14pt}
\hoffset-23.5pt


\title{Triplet Cooper Pair Formation by 
Anomalous Spin Fluctuations\\
in Non-centrosymmetric Superconductors}

\author{Tetsuya Takimoto$^{1,2}$ and Peter Thalmeier$^1$}

\inst{
$^1$Max Planck Institute for Chemical Physics of Solids, 
N\"{o}thnitzer Str. 40, 01187 Dresden, Germany\\
$^2$Asia Pacific Center for Theoretical Physics, 
Pohang University of Science and Technology, 
Pohang, Gyeongbuk 790-784, Korea
}

\recdate{\today}

\abst{
A microscopic theory for the spin triplet Cooper pairing 
in non-centrosymmetric superconductors like 
CePt$_3$Si and CeTSi$_3$ (T=Rh, Ir) 
is presented. The lack of inversion symmetry leads to new 
anomalous spin fluctuations which stabilize the triplet part 
in addition to the singlet part originating 
from the centrosymmetric spin fluctuations. 
It is shown that both parts have similar nontrivial momentum dependence 
of A$_1$ type. 
Therefore the mixed singlet-triplet gap function has accidental line nodes 
on both Fermi surface sheets which are stable as function of temperature. 
This gap function explains the salient features of CePt$_3$Si and CeTSi$_3$ 
superconductors. 
}
\kword{
non-centrosymmetric system, antisymmetric spin-orbit coupling, 
superconductivity}

\maketitle

Since its discovery in the heavy fermion compound 
CeCu$_2$Si$_2$\cite{CeCu2Si2}, 
unconventional superconductivity has attracted much attention 
in the field of correlated electron systems. 
For systems with inversion symmetry, 
the pairing states of superconductivity are classified as 
spin-singlet of even-parity or spin-triplet of odd-parity. 
Recalling that a phonon-mediated mechanism is established for 
s-wave spin-singlet superconductivity, 
the possibility of spin-triplet superconductivity, 
as realized, e.g. in UPt$_3$ and Sr$_2$RuO$_4$, 
is considered as a characteristic 
feature of the correlated electron system. 

It has been suggested that the spin-triplet superconductivity 
is destroyed by breaking the inversion symmetry
\cite{Anderson}. 
However, recently, the unconventional superconductivity has been observed 
in several non-centrosymmetric heavy fermion compounds like 
CePt$_3$Si\cite{CePt3Si}, CeRhSi$_3$\cite{CeRhSi3}, 
CeIrSi$_3$\cite{CeIrSi3}, and CeCoGe$_3$\cite{CeCoGe3}. 
These compounds have tetragonal crystal structures without 
the inversion symmetry, and the superconductivity is found 
around each antiferromagnetic phase
\cite{Tateiwa,CeRhSi3,CeIrSi3,CeCoGe3}. 
The evidence of unconventional superconductivity is at least 
shown by the existence of a line node of the superconducting gap 
in CePt$_3$Si, which is observed by 
NMR\cite{Yogi}, thermal transport\cite{Izawa}, 
and penetration depth\cite{Bonalde}. 
Furthermore, the common remarkable property 
of the unconventional superconductivity 
is the large upper critical magnetic field 
exceeding the Pauli limiting field\cite{CePt3Si,CeRhSi3,CeIrSi3,Settai}. 
The observed upper critical field has opened up 
the possibility of the spin-triplet superconductivity 
in non-centrosymmetric systems in contrast to the theoretical expectation
\cite{Anderson}. 

The lack of inversion symmetry of non-centrosymmetric systems 
leads to an antisymmetric spin-orbit interaction
\cite{Rashba}. 
A general superconducting order parameter 
is described by the even spin-singlet gap function $\psi({\bf k})$ 
and odd spin-triplet gap function ${\bf d}({\bf k})$\cite{SigristUeda}. 
The coexistence of $\psi({\bf k})$ with ${\bf d}({\bf k})$ 
is unavoidable in principle in the non-centrosymmetric superconductor 
because of the lack of inversion symmetry. 
For the non-centrosymmetric superconductor with a large upper critical 
field, Sigrist {\it et al.} have reconsidered 
the possibility of spin-triplet pairing and concluded that 
a Cooper pair satisfying 
${\bf d}({\bf k})\parallel{\bf g}_{\bf k}$ (${\bf g}_{\bf k}$: Rashba field) 
does not suffer the suppression of pairing coming from the lack of 
inversion symmetry\cite{Frigeri1}. 
Based on this spin-triplet state, many theoretical proposals 
have been made for the superconducting properties. 
However, the pairing mechanism remains 
an unsettled problem except for a few attempts
\cite{Yokoyama,Tada,Yanase}. 

On the other hand, a spin fluctuation property characteristic 
of the non-centrosymmetric system has been developed recently
\cite{Takimoto}. 
Unlike the centrosymmetric system, the anomalous spin fluctuations 
do not vanish in the non-centrosymmetric system under the condition, 
that the symmetry of the momentum dependence is equivalent to 
the symmetry of spin-product included in the spin fluctuation. 
Furthermore, not only usual (centrosymmetric) spin fluctuations 
but also anomalous (non-centrosymmetric) spin fluctuations enhance 
on approaching the magnetic instability. 
Therefore, it is expected that the latter play an essential role 
in the pairing mechanism of superconductivity. 

In this Letter, 
we study the superconducting transition in a correlated electron 
system without inversion symmetry. 
The roles of usual and anomalous spin fluctuations 
in the superconductivity and the nodal structure of gap functions 
are examined within the weak-coupling theory. 
We also compare the superconductivity of centrosymmetric systems 
with that of non-centrosymmetric ones around the magnetic instability.


The latter is described by 
the following Hamiltonian $H=H_0+H_1$ for correlated electrons:
\begin{eqnarray}
  &&H_0=\sum_{{\bf k}\sigma\sigma'}
         [(\varepsilon_{\bf k}-\mu)\hat{\sigma}_{0}
          +{\bf g}_{\bf k}\cdot\hat{{\bf \sigma}}]_{\sigma\sigma'}
              c_{{\bf k}\sigma}^{\dagger}c_{{\bf k}\sigma'},\\
  &&H_1=U\sum_{\bf i}n_{{\bf i}\uparrow}n_{{\bf i}\downarrow},
\end{eqnarray}
where $c_{{\bf k}\sigma}$ and $c_{{\bf k}\sigma}^{\dagger}$ are 
annihilation and creation operators of an electron 
with a momentum ${\bf k}$ and a spin $\sigma$. Here, $\varepsilon_{\bf k}$ 
and $\mu$ are the energy dispersion of electrons and the chemical potential, 
respectively, 
while ${\bf g}_{\bf k}$=$-{\bf g}_{\bf -k}$ describes the Rashba field 
coming from the antisymmetric spin-orbit interaction, 
which breaks the inversion symmetry. 
Then, eigenenergies of $H_0$ are given by 
$\varepsilon_{{\bf k}\pm}=\varepsilon_{\bf k}\pm|{\bf g}_{\bf k}|-\mu$. 
In $H_1$, $U$ is the screened on-site interaction. 

In the following, we consider a two-dimensional tetragonal system 
with a dispersion energy 
$\varepsilon_{\bf k}=2t_1(\cos{k_x}+\cos{k_y})+4t_2\cos{k_x}\cos{k_y}$ 
and a Rashba-field ${\bf g}_{\bf k}=g(\sin{k_y},-\sin{k_x},0)$, 
which is a periodic form of the simplest one ($k_y$, -$k_x$, 0)
\cite{Frigeri1}. 
Chosing parameters as $t_2/t_1$=0.35 and $g/t_1$=0.2, 
we can reproduce quasi-two dimensional Fermi surfaces of CePt$_3$Si 
obtained by band calculations\cite{Samokhin1,Hashimoto}. 
Furthermore, the reason why the two dimensional system is used 
instead of the realistic three dimensional system is based on 
the following knowledge. 
It is known that the unconventional superconducting phase shrinks with 
increasing the dimensionality in the centrosymmetric system
\cite{TM}. 
Recalling that the superconducting transition temperature 1.6 K
in a non-centrosymmetric system CeIrSi$_3$ is in the same order as 
the highest superconducting transition temperature 2.7 K of CeCoIn$_5$ 
among Ce-based compounds\cite{Petrovic}, 
it is suggestive that unconvensional superconductivity is 
more favorable 
in two dimensional system than in three dimensional one, 
independent of the inversion symmetry. 

From the specific heat data\cite{Tateiwa2} and $2\Delta/T_{\rm c}$ value
used to fitting to the NMR relaxation rate\cite{Mukuda} 
in CeIrSi$_3$, the strong-coupling theory is suggested. 
However, the weak-coupling theory will be sufficient to examine 
the mechanism of superconductivity. 
In general, 
the matrix gap function $\hat{\Delta}({\bf k})$ 
is decomposed 
into a spin-singlet $\psi({\bf k})$ and a spin-triplet 
${\bf d}({\bf k})$-vector as 
$
 \hat{\Delta}({\bf k})=[\psi({\bf k})\hat{\sigma}_0
 +{\bf d}({\bf k})\cdot\hat{\bf \sigma}]{\rm i}\hat{\sigma}_y
$
\cite{SigristUeda}. 
In non-centrosymmetric superconductors, 
only spin-triplet component satisfying 
$|{\bf d}({\bf k})\cdot{\bf g}_{\bf k}|$
=$|{\bf d}({\bf k})||{\bf g}_{\bf k}|$ 
is not affected by the suppression of pairing coming from the lack of 
inversion symmetry\cite{Frigeri1}. 
Then, the spin-triplet component will be given by 
${\bf d}({\bf k})$=$\phi({\bf k}){\bf g}_{\bf k}$, 
where the symmetry of momentum dependence of $\phi({\bf k})$ 
is the same as that of the spin-singlet $\psi({\bf k})$
\cite{Fujimoto1,Tada}. 
Under this condition for ${\bf d}({\bf k})$, 
the normal and anomalous matrix Green's functions, 
$\hat{G}({\bf k}, {\rm i}\omega_n)$ and 
$\hat{F}({\bf k}, {\rm i}\omega_n)$, respectively, 
are defined 
with $\tilde{\bf g}_{\bf k}$=${\bf g}_{\bf k}/|{\bf g}_{\bf k}|$ as
\begin{eqnarray}
 &&\hat{G}({\bf k}, {\rm i}\omega_n)
   =G_+({\bf k}, {\rm i}\omega_n)\hat{\sigma}_0
   +G_-({\bf k}, {\rm i}\omega_n)\tilde{\bf g}_{\bf k}\cdot\hat{\bf \sigma},\\
 &&\hat{F}({\bf k}, {\rm i}\omega_n)
   =[F_+({\bf k}, {\rm i}\omega_n)\hat{\sigma}_0
    +F_-({\bf k}, {\rm i}\omega_n)\tilde{\bf g}_{\bf k}\cdot\hat{\bf \sigma}]
             {\rm i}\hat{\sigma}_y,
\end{eqnarray}
$G_{\pm}({\bf k}, {\rm i}\omega_n)$ and 
$F_{\pm}({\bf k}, {\rm i}\omega_n)$ are given by
\begin{eqnarray}
 &&G_{\pm}({\bf k}, {\rm i}\omega_n)
             =\frac{1}{2}\left[\frac{-{\rm i}\omega_n-\varepsilon_{{\bf k}+}}
                               {\omega_n^2+E_{{\bf k}+}^2}
                       \pm\frac{-{\rm i}\omega_n-\varepsilon_{{\bf k}-}}
                               {\omega_n^2+E_{{\bf k}-}^2}\right],\\
 &&F_{\pm}({\bf k}, {\rm i}\omega_n)
             =\frac{1}{2}\left[\frac{\Delta_{{\bf k}+}}
                               {\omega_n^2+E_{{\bf k}+}^2}
                       \pm\frac{\Delta_{{\bf k}-}}
                               {\omega_n^2+E_{{\bf k}-}^2}\right],
\end{eqnarray}
with $\Delta_{{\bf k}\pm}$=$\psi({\bf k})\pm\phi({\bf k})|{\bf g}_{\bf k}|$ 
and $E_{{\bf k}\pm}$=$\sqrt{\varepsilon_{{\bf k}\pm}^2+\Delta_{{\bf k}\pm}^2}$.


Within the weak-coupling theory for superconductivity, 
only the static spin susceptibility is required. 
For simplicity, we neglect the feedback effect of superconductivity 
on the spin fluctuation below $T_{\rm c}$. 
Therefore, in this case the spin fluctuation is just 
the static spin susceptibility in the normal state. 
Including the electron repulsion $U$ within RPA, 
the matrix of static spin susceptibility is calculated 
with the above Green's function 
as 
$
 \hat{\chi}({\bf q})=
 \left[{\bf 1}-2U\hat{\chi}^{(0)}({\bf q})\right]^{-1}
 \hat{\chi}^{(0)}({\bf q}),
\label{chirpa}
$
where the matrix element $\chi_{\alpha\beta}^{(0)}({\bf q})$ 
for $\alpha, \beta$=$x, y, z$ is given by
\begin{eqnarray}
 \chi^{(0)}_{\alpha\beta}({\bf q})
 =\frac{1}{8N_0}\sum_{\bf k}\sum_{\xi,\zeta}
  \Gamma_{\xi\zeta}^{\alpha\beta}({\bf k}; {\bf q})
   \frac{f(\varepsilon_{{\bf k}\zeta})-f(\varepsilon_{{\bf k+q}\xi})}
        {\varepsilon_{{\bf k+q}\xi}-\varepsilon_{{\bf k}\zeta}},
\end{eqnarray}
with the Fermi distribution function $f(\varepsilon)$ and a vertex
\begin{eqnarray}
 &&\Gamma_{\xi\zeta}^{\alpha\beta}({\bf k}; {\bf q})=
  \delta_{\alpha,\beta}
   \left(1-\xi\zeta\tilde{\bf g}_{\bf k}\cdot\tilde{\bf g}_{\bf k+q}\right)
\\
  &&+\xi\zeta\left(\tilde{g}_{{\bf k}\alpha}\tilde{g}_{{\bf k+q}\beta}
                +\tilde{g}_{{\bf k}\beta}\tilde{g}_{{\bf k+q}\alpha}\right)
  -\epsilon_{\alpha\beta\gamma}{\rm i}
        \left(\xi\tilde{g}_{{\bf k+q}\gamma}
             -\zeta\tilde{g}_{{\bf k}\gamma}\right).\nonumber
\end{eqnarray}
Similarly, the charge fluctuation is described by 
$\chi_{cc}({\bf q})=\chi^{(0)}_{cc}({\bf q})/[1+2U\chi^{(0)}_{cc}({\bf q})]$ 
with $\chi^{(0)}_{cc}({\bf q})$, which is obtained by the replacement 
of $\Gamma_{\xi\zeta}^{\alpha\beta}({\bf k}; {\bf q})$ 
by $\Gamma_{\xi\zeta}^{cc}({\bf k}; {\bf q})$=
$1+\xi\zeta\tilde{\bf g}_{\bf k}\cdot\tilde{\bf g}_{\bf k+q}$ 
in $\chi^{(0)}_{\alpha\beta}({\bf q})$. 

We comment on the spin fluctuations in the tetragonal system. 
In the centrosymmetric system, the matrix $\hat{\chi}({\bf q})$ 
is diagonal. We denote 
$\chi_{zz}({\bf q})$ and 
$(\chi_{xx}({\bf q})+\chi_{yy}({\bf q}))/2$ 
as usual (centrosymmetric) spin fluctuations. 
On the other hand, in the non-centrosymmetric case, 
$(\chi_{xx}({\bf q})-\chi_{yy}({\bf q}))/2$, 
$(\chi_{xy}({\bf q})+\chi_{yx}({\bf q}))/2$, 
$(\chi_{yz}({\bf q})-\chi_{zy}({\bf q}))/2{\rm i}$, and 
$(\chi_{zx}({\bf q})-\chi_{xz}({\bf q}))/2{\rm i}$ also remain non-zero 
with characteristic ${\bf q}$-dependences 
of $q_x^2-q_y^2$-, $q_xq_y$-, $q_y$-, and $q_x$-type, respectively 
\cite{Takimoto}. 
We call these four contributions as anomalous spin fluctuations. 
The characteristic ${\bf q}$-dependences of anomalous 
(non-centrosymmetric) spin fluctuations 
are caused by the Rashba field ${\bf g}_{\bf k}$ 
under the symmetry constraint. 
Considering the matrix $\hat{\chi}({\bf q})$, 
not only usual spin fluctuations but also anomalous spin fluctuations 
develop around a magnetic instability. 


We now examine the effect of both usual and anomalous 
spin fluctuations on the pairing mechanism 
in the non-centrosymmetric tetragonal system. 
Using the standard procedure\cite{AB,Nakajima,Miyake,Scalapino},
the following gap equation is obtained,
\begin{eqnarray}
  &&\left[
   \begin{array}{ccc}
    \psi({\bf k}) & d_x({\bf k}) & d_y({\bf k})
   \end{array}
  \right]^t\nonumber\\
  &=&\frac{1}{N_0}\sum_{\bf q}
  \left[
   \begin{array}{ccc}
    V_{ss}({\bf q}) & V_{sx}({\bf q}) & V_{sy}({\bf q})\\
    V_{xs}({\bf q}) & V_{xx}({\bf q}) & V_{xy}({\bf q})\\
    V_{ys}({\bf q}) & V_{yx}({\bf q}) & V_{yy}({\bf q})\\
    \end{array}
  \right]
  \left[
   \begin{array}{c}
    F_s({\bf k-q})\\
    F_x({\bf k-q})\\
    F_y({\bf k-q})\\
   \end{array}
  \right],
\end{eqnarray}
where $V_{\xi\zeta}({\bf q})$ is the pairing interaction 
due to corresponding fluctuation exchange given by
\begin{eqnarray}
 V_{ss}({\bf q})&=&-U^2
   [\chi_{zz}({\bf q})+\chi_{xx}({\bf q})
   +\chi_{yy}({\bf q})-\chi_{cc}({\bf q})]-U\label{vss}\nonumber\\
 V_{xx}({\bf q})&=&U^2
   [\chi_{cc}({\bf q})+\chi_{zz}({\bf q})
   -\{\chi_{xx}({\bf q})-\chi_{yy}({\bf q})\}],\nonumber\\
 V_{yy}({\bf q})&=&U^2
   [\chi_{cc}({\bf q})+\chi_{zz}({\bf q})
   +\{\chi_{xx}({\bf q})-\chi_{yy}({\bf q})\}],\nonumber\\
 V_{xy}({\bf q})&=&V_{yx}({\bf q})
  =-U^2
   [\chi_{xy}({\bf q})+\chi_{yx}({\bf q})],\\
 V_{sx}({\bf q})&=&-V_{xs}({\bf q})
  =i U^2
   [\chi_{yz}({\bf q})-\chi_{zy}({\bf q})],\nonumber\\
 V_{sy}({\bf q})&=&-V_{ys}({\bf q})
  =i U^2
   [\chi_{zx}({\bf q})-\chi_{xz}({\bf q})].\nonumber
\end{eqnarray}
Similar relations between the pairing interactions and spin fluctuations 
in non-centrosymmetric systems 
are obtaind from a different approach\cite{Samokhin2}. 
In order to get a favorable form of the gap equation, 
$F_s({\bf k})$ and $F_{\alpha}({\bf k})$ 
are introduced by
$
 T\sum_{n}\hat{F}({\bf k},{\rm i}\omega_n)
 =[F_s({\bf k})\hat{\sigma}_0+{\bf F}({\bf k})\cdot\hat{\bf \sigma}]
  {\rm i}\hat{\sigma}_y,
$
where $F_s({\bf k})$=$[\psi({\bf k})\varphi_+({\bf k})
+{\bf d}({\bf k})\cdot\tilde{\bf g}_{\bf k}\varphi_-({\bf k})]/2$ 
and $F_{\alpha}({\bf k})$=
$\tilde{g}_{{\bf k}\alpha}[\psi({\bf k})\varphi_-({\bf k})
+{\bf d}({\bf k})\cdot\tilde{\bf g}_{\bf k}\varphi_+({\bf k})]/2$ 
are obtained with 
$\varphi_{\pm}({\bf k})$=$\tanh{\frac{E_{{\bf k}+}}{2T}}/2E_{{\bf k}+}
\pm\tanh{\frac{E_{{\bf k}-}}{2T}}/2E_{{\bf k}-}$. 

In the centrosymmetric case, only $V_{ss}({\bf q})$ and 
$V_{xx}({\bf q})=V_{yy}({\bf q})$ remain, 
and the usual spin fluctuations in $V_{ss}({\bf q})$ and $V_{xx}({\bf q})$ 
contribute to the spin-singlet and spin-triplet pairing mechanisms, 
respectively. 
For non-zero Rashba field, 
the anomalous spin fluctuations contribute {\it only} 
to the spin-triplet pairing interactions $V_{\alpha\beta}({\bf q})$ and 
the mixing interactions $V_{s\alpha}({\bf q})=-V_{\alpha s}({\bf q})$ 
for $\alpha$,$\beta$=x, y. We also note that 
the mixing interactions 
between spin-singlet and spin-triplet pairs are described by 
antisymmetric spin fluctuations, $\chi_{yz}({\bf q})-\chi_{zy}({\bf q})$ 
and $\chi_{zx}({\bf q})-\chi_{xz}({\bf q})$, 
which relate with 
the Dzyaloshinski-Moriya interaction\cite{Frigeri2}. 
Therefore, the anomalous spin fluctuations are essential to form 
the spin-triplet pairs in the non-centrosymmetric system, 
if they are constructive for superconductivity.

\begin{figure}[t]
\begin{center}
\resizebox{60mm}{!}
{\includegraphics{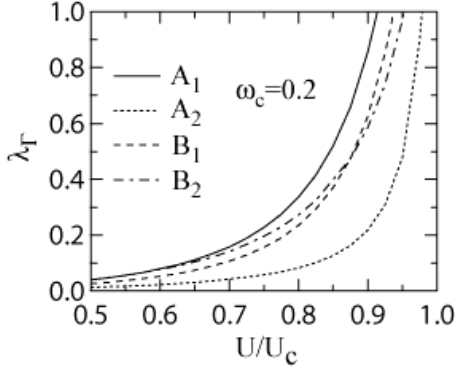}}
\end{center}
\caption{Interaction dependence of maximum eigenvalue 
for superconductivity belonging to $\Gamma$-representation 
in $C_{4v}$ point group of the non-centrosymmetric system.}
\end{figure}

In the following, we show the results of numerical calculation. 
When the system is in the normal state, 
the gap equation reduces to an eigenvalue problem. 
When the maximum eigenvalue of a representation reaches unity, 
superconductivity belonging to this representation appears. 
For actual calculation, we introduce a cutoff energy 
$\omega_{\rm c}$=0.2$t_1$ for electrons forming the Cooper pair. 
Furthermore, 
we fix the superconducting transition temperature 
at $T_{\rm c}$=0.02$t_1$. 
In Fig. 1, it is shown that by increasing the on-site repulsion $U$ 
toward an incommensurate magnetic instability at $U_{\rm c}$=2.551$t_1$
\cite{Takimoto}, 
every maximum eigenvalue $\lambda_{\Gamma}$ of $\Gamma$-representation 
for the eigenvalue problem increases, and $\lambda_{A_1}$ reaches 
unity first among all tetragonal group representations. 
This means that the superconductivity of $A_1$ representation 
in $C_{4v}$ appears around the incommensurate magnetic instability.

\begin{figure}[t]
\begin{center}
\resizebox{85mm}{!}
{\includegraphics{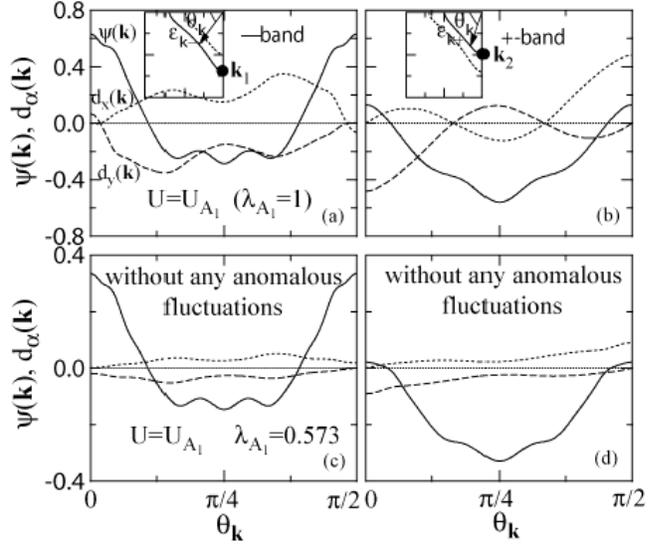}}
\end{center}
\caption{(a) and (b): ${\bf k}$-dependence of gap functions, 
$\psi({\bf k})$ (solid), $d_x({\bf k})$ (dotted), and $d_y({\bf k})$ (dashed) 
along Fermi-surfaces of $\varepsilon_{{\bf k}\pm}$ 
for the full gap equation 
with the critical interaction constant 
satisfying $\lambda_{A_1}$=1. The insets show 
Fermi-surfaces in the 1$^{\rm st}$-quadrant of Brillouin zone. 
(c) and (d): ${\bf k}$-dependence of gap functions 
along Fermi-surfaces of $\varepsilon_{{\bf k}\pm}$ 
for a gap equation without any anomalous spin fluctuations 
using the same interaction constant in (a) and (b). }
\end{figure}

For the eigenvector of $\lambda_{A_1}$=1 
at the critical interaction constant $U_{A_1}$, 
${\bf k}$-dependences of the corresponding normalized gap functions 
along Fermi-surfaces of $\varepsilon_{{\bf k}-}$ and 
$\varepsilon_{{\bf k}+}$ are shown in Figs. 2(a) and 2(b), respectively. 
Ignoring all anomalous spin fluctuations with $U$=$U_{A_1}$ 
in the gap equation, the ${\bf k}$-dependences of 
the gap functions change to Figs. 2(c) and 2(d), respectively, 
where $\lambda_{A_1}$ decreases to 0.573 although 
the same interaction constant as Figs. 2(a) and 2(b) are used. 
Comparing these figures, the singlet gap function $\psi({\bf k})$ 
is not affected by the anomalous spin fluctuations, 
as expected from eq. (\ref{vss}). 
Therefore, it is stabilized by the usual spin 
fluctuations with a momentum corresponding to the ordering wave vector 
at $U_{\rm c}$, which spans from one peak position of $\psi({\bf k})$ 
along a Fermi-line $\varepsilon_{{\bf k}-}$=0 in Fig. 2(a) 
to other peak position of $|\psi({\bf k})|$ 
along a Fermi-line $\varepsilon_{{\bf k}+}$=0 in Fig. 2(b)
\cite{Takimoto}. 
On the other hand, the magnitude of spin-triplet gap functions 
are enhanced by switching on the anomalous spin fluctuations, 
and becomes almost the same size as that of spin-singlet gap. 
Thus, the anomalous spin fluctuations are constructive 
for the superconductivity, 
and they are responsible for the stabilization of 
spin-triplet gap function. 
We note that the effect of anomalous spin fluctuations on superconductivity 
does not depend on $\omega_{\rm c}$.

\begin{figure}[t]
\begin{center}
\resizebox{85mm}{!}
{\includegraphics{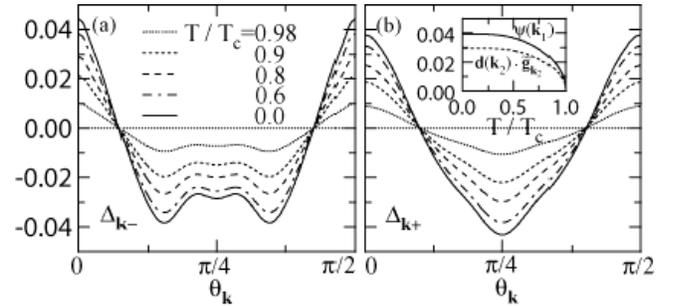}}
\end{center}
\caption{${\bf k}$-dependence of gap functions $\Delta_{{\bf k}\pm}$ 
along Fermi-surface of $\varepsilon_{{\bf k}\pm}$ 
at several temperatures below $T_{\rm c}$. Inset shows 
the temperature dependences of spin-singlet and spin-triplet 
gap functions at momenta on Fermi-surfaces giving the corresponding 
maximum values of 
$\psi({\bf k}_1)$ (solid) and 
${\bf d}({\bf k}_2)\cdot\tilde{\bf g}_{{\bf k}_2}$ (dashed), 
where ${\bf k}_{\rm i}$ are shown in insets of Fig. 2.}
\end{figure}

Due to the nodal structures and similar amplitudes of 
$\psi({\bf k})$ and ${\bf d}({\bf k})$, 
nodal structures for $\Delta_{{\bf k}\pm}$ are also expected. 
Considering that $\Delta_{{\bf k}\pm}$ are the gap functions 
in the band picture, they reflect directly to 
thermal and dynamical quantities in 
the superconducting state. 
For various temperatures below $T_{\rm c}$=0.02$t_1$, 
the ${\bf k}$-dependences of 
$\Delta_{{\bf k}\pm}$ along Fermi surfaces of $\varepsilon_{{\bf k}\pm}$ 
are shown in Fig. 3, 
where pairing interactions are fixed to those at $T=T_{\rm c}$. 
We stress that the ${\bf k}$-dependences of gap functions reflect fully 
the characteristic momentum dependences of spin fluctuations 
without assuming the common simple form of 
$\Delta_{{\bf k}\pm}=\Delta_{s}\pm\Delta_{t}|{\bf g}_{\bf k}|$. 
Recalling that the representation of the superconductivity is 
$A_1$ in $C_{4v}$, 
{\it both} gap functions $\Delta_{{\bf k}\pm}$ 
exhibit accidental line nodes. 
The positions of nodes almost do not move with decreasing temperature, 
and the temperature dependence of 
$\psi({\bf k}_1)$ and ${\bf d}({\bf k}_2)\cdot\tilde{\bf g}_{{\bf k}_2}$ 
is of common BCS type as shown in the inset. 
Therefore, the existence of the accidental line node in gap function 
explains $T$-dependences of NMR relaxation rate, thermal conductivity, 
and penetration depth in the superconducting state of CePt$_3$Si 
and CeIrSi$_3$
\cite{Yogi,Izawa,Bonalde,Mukuda}. 
In order to check our scenario, 
it is desirable to observe the gap function in detail by 
the thermal conductivity experiment, in addition to the observation of 
anomalous spin fluctuations. 

Finally, we comment on the superconductivity 
in non-centrosymmetric correlated electron systems. 
As already mentioned, the spin-singlet and spin-triplet gap functions 
coexist in the non-centrosymmetric superconductor. 
If only the usual antiferromagnetic spin fluctuations are enhanced, 
the spin-triplet gap function will be induced by the primary order 
parameter corresponding to the spin-singlet gap function, 
as shown in Figs. 2(c) and 2(d)\cite{Tada}. 
However, in the case that the anomalous spin fluctuations 
are also enhanced\cite{Takimoto}, the superconducting state 
is quite different from the former case, 
although it will not be so different 
from that of the usual unconventional line-node superconductivity. 
With respect to its relation to magnetism, we note that 
unconventional superconductivity appears due to the spin fluctuations 
enhanced around the magnetic instability 
in a similar manner as in a centrosymmetric superconductor. 
On the other hand, the spin-triplet gap function develops due to 
the anomalous spin fluctuations characteristic of 
the non-centrosymmetric structure. 

In summary, we have studied the mechanism of 
superconductivity in the non-centrosymmetric system. 
It has been shown that 
the spin-triplet gap function surviving the suppression of pairing 
due to the absence of centrosymmetry is developed by the anomalous spin 
fluctuations, which are enhanced around a magnetic instability. 
Reflecting the anisotropic momentum dependences of anomalous spin 
fluctuations, the gap functions have non-trivial A$_1$ 
${\bf k}$-dependence with accidental line node. 
This pairing mechanism gives a reasonable explanation 
for the superconductivity in a non-centrosymmetric superconducting systems 
CePt$_3$Si and CeTSi$_3$, 
showing a huge upper critical field, a line-node gap structure, 
and an incommensurate magnetic structure. 


\vskip-0.5cm

\end{document}